# Towards Semi-Markov Model-based Dependability Evaluation of VM-based Multi-Domain Service Function Chain

Lina Liu, Jing Bai, Xiaolin Chang, Fumio Machida, Kishor S. Trivedi, Haoran Zhu

**Abstract—**In NFV networks, service functions (SFs) can be deployed on virtual machines (VMs) across multiple domains and then form a service function chain (MSFC) for end-to-end network service provision. However, any software component in a VM-based MSFC must experience software aging issue after a long period of operation. This paper quantitatively investigates the capability of proactive rejuvenation techniques in reducing the damage of software aging on a VM-based MSFC. We develop a semi-Markov model to capture the behaviors of SFs, VMs and virtual machine monitors (VMMs) from software aging to recovery under the condition that failure times and recovery times follow general distributions. We derive the formulas for calculating the steady-state availability and reliability of the VM-based MSFC composed of multiple SFs running on VMs hosted by VMMs. Sensitivity analysis is also conducted to identify potential dependability bottlenecks.

## 1 Introduction

NETWORK Function Virtualization (NFV) dissociates network functions from the underlying hardware, enabling the agility and flexibility of deploying new network services to support growing customer demands [1]. In NFV networks, service functions (SFs) deployed on virtual machines (VMs) are usually chained together to form a service function chain (SFC) [2]. It is a fact that some domain network resources can be only accessed by specific SFs [3]. Therefore, multiple SFs forming an SFC may span multi-domain networks (as shown in Fig. 1) for providing end-to-end network services [4]. Applications of SFCs in multiple domains (MSFCs) include consumer broadband, mobile packet core and mobile backhaul [5].

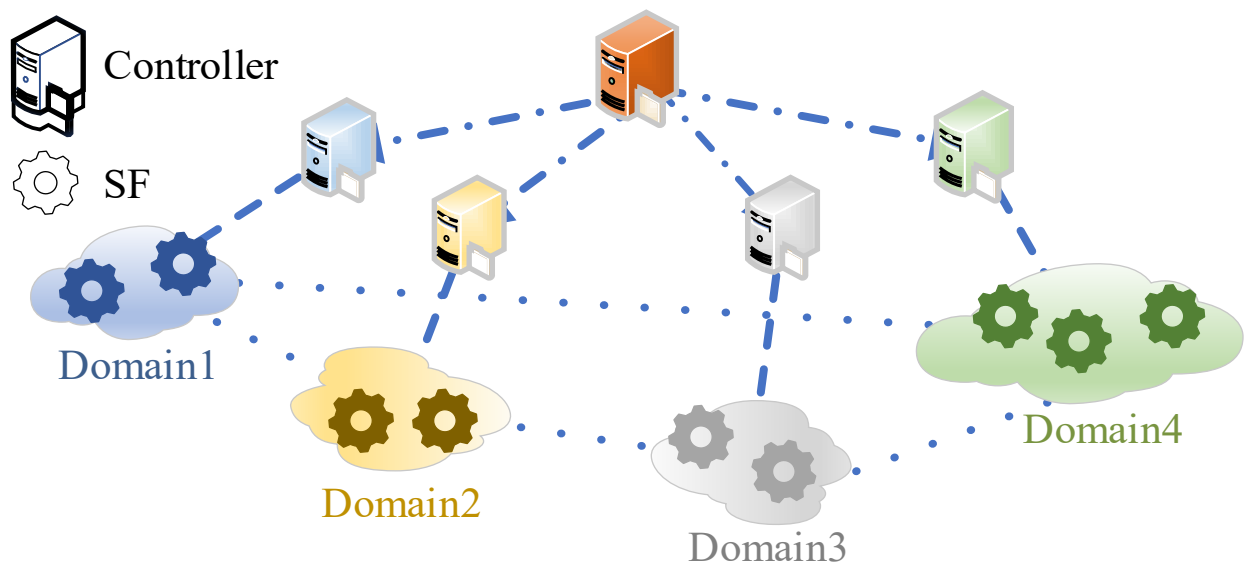

**Fig. 1** MSFC architecture

In a VM-based MSFC, all SFs, VMs and VMMs running are subject to software aging [6], leading to the decrease in the MSFC dependability (such as the availability and reliability) [7]-[9]. Proactive rejuvenation techniques can reduce this decrease of the dependability of a VM-based MSFC [10]. To evaluate the effectiveness of proactive rejuvenation techniques, several analytical models have been used to analyze the dependability of various systems, such as virtualization systems [13]-[31]. However, the models developed in [13]-[18] only involved a single SF but cannot be directly used to capture the behaviors of SFC systems with multiple SFs. While the models developed in [19]-[31] analyzed SFC system behaviors, they cannot capture the dynamic interaction among SFs, VMs and VMMs in the VM-based MSFC. The following lists two major challenges to be solved in modeling the VM-based MSFC.

- The dependency between the upper and lower components in the VM-based MSFC is more complex than that in the container-based MSFC, as illustrated by the red lines in Fig. 2. When the VMM suffers from failure due to software aging, all SFs and VMs running on this VMM become unavailable,which leads to a decrease in VM-based MSFC dependability. In addition, each component has different resource requirements due to the different service requirements. This results in different time intervals of abnormal events and recovery events for each component. Thus, capturing the differentiated behaviors between SFs and VMs, between SFs and VMMs, and between VMs and VMMs in each host of the VM-based MSFC is a challenge.
- Multiple time-dependent interactions between components in the VM-based MSFC are more complex than that in the container-based MSFC, as shown by the green lines in Fig. 2. That is, there are dependencies between multiple SFs, between multiple VMs, between multiple VMMs, between VMs and SFs deployed on other VMs, between VMMs and SFs deployed on other VMMs, and between VMMs and VMs deployed on other VMMs. Thus, capturing multiple time-dependent interactions between components in the VM-based MSFC is a challenge.

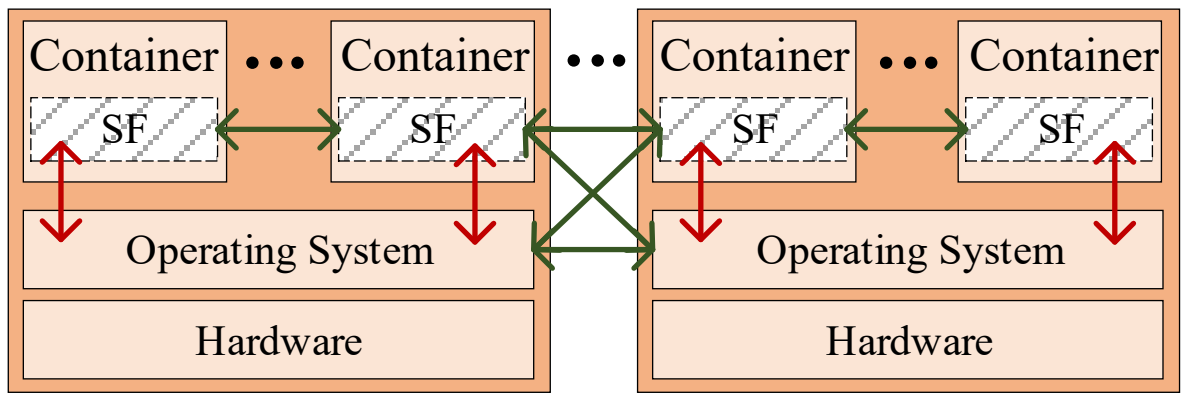

(a) Container-based MSFC

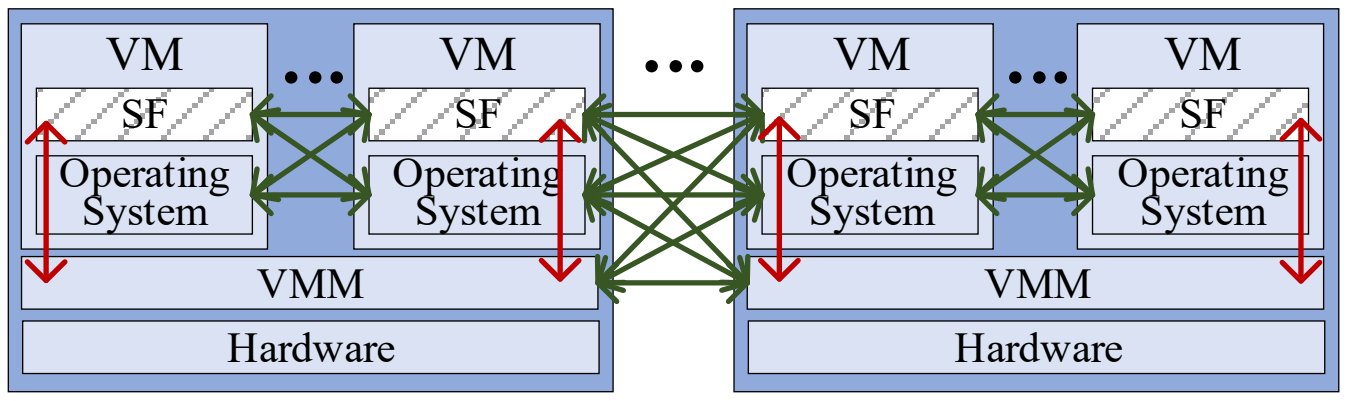

(b) VM-based MSFC

**Fig. 2** Comparison of a container-based MSFC and VM-based MSFC

To address the above challenges, we develop a semi-Markov process (SMP) model to quantitatively analyze the service dependability in the VM-based MSFC system which deploys proactive rejuvenation techniques against software aging. To the best of our knowledge, we are the first to develop a monolithic SMP model of a VM-based SFC. The contributions of this paper are summarized as follows:

- A novel comprehensive SMP model is proposed to quantitatively evaluate the effectiveness of proactive rejuvenation techniques. This model describes the behaviors of VM-based MSFC using proactive rejuvenation techniques from software aging to recovery. In this model, the failure time and recovery time follow the general distribution. In addition, our model can capture the time-dependent behaviors between the components (SFs, VMs and VMMs) in a VM-based MSFC.
- We not only evaluate the steady-state availability of VM-based MSFC, but also consider the mean time to failure (MTTF) as a reliability metric. Specifically, we derive the formulas of calculating these two metrics of a VM-based MSFC composed of any number of SFs running on any number of VMs hosted by any number of VMMs.
- We demonstrate the approximation accuracy and effectiveness of the proposed model and metric formulas by simulation and numerical experiments, respectively. Numerical experiments show that proactive rejuvenation technique can effectively counteract software aging in VM-based MSFC. The sensitivity analysis shows that the aging time is an important parameter for the VM-based MSFC dependability. Moreover, when both the failure time and recovery time follow non-exponential distributions, there is a significant difference in the results of availability and reliability compared to when all event times follow exponential distributions. Through quantitative analysis, the service provider can dynamically adjust the system parameters to maximize their benefit.

The rest of this paper is set up as follows. Section 2 introduces the related work. The SMP model and metric formulas are given in Section 3 and Section 4, respectively. The results of numerical and simulation experiments are given in Section 5. The conclusion is given in Section 6.

## 2 Related Work

Diverse analytical models have been proposed to quantify the dependability attributes. Analytical models for dependability evaluation are typically classified into three categories: non-state-space, state-space and hierarchical models. Non-state space models, such as the reliability block diagram (RBD), reliability graph (RG), and fault-tree (FT), assume that component failure and repair behaviors are independent and do not allow time-dependence [11]. State-space models, such as the continuous-time Markov chain (CTMC) and SMP, can model time-dependent interactions and behaviors within a system [12]. In hierarchical modeling approaches are multi-layer models, in which the higher layer usually applies a non-state-space model to represent the relationships of individual subsystems, while the lower layer usually applies a state-space model that is better suited to capture individual complex behaviors [12]. These analytical models have been used to evaluate the dependability of non-SFC systems [13]-[18] and SFC system [19]-[31].

### 2.1 Non-SFC System Modeling Approaches

Many analytical models have been proposed to evaluate the availability and reliability of virtualization systems.

i). *State-space model*. Based on the state-space modeling approach, the authors in [13] constructed a CTMC model to capture the behavior of the edge user plane function (EUPF) service system with the deployment of rejuvenation techniques and perform transient and steady-state analysis of EUPF service dependability. In [14], Martyna established a SMP model to predict the availability and reliability of mobile cloud computing (MCC) systems. Based on the Markov regenerative process (MRGP) model, the authors in [15] captured the behaviors of the system consisting of application service, VM, and VMM. The experiment injecting memory leaks was carried out to measure the software aging related parameters. In [16], Zhang *et al.* proposed an availability model of the virtualization system with the deployment of live migration technique, and performed numerical analysis experiments to quantify the impact of parameter uncertainty on the steady-state availability of the system. In [17], Torquato *et al.* evaluated the performability and dependability of five virtualized environments covering different bursty workload conditions.

ii). *Other model*. In [18], Clemente *et al.* proposed a hierarchical availability modeling method based on RBD and stochastic Petri net to analyze redundant cloud systems.

The above models [13]-[18] involved a single SF, and can not directly capture the behaviors of SFC with multiple SFs.

### 2.2 SFC System Modeling Approaches

Our work focuses on evaluating the dependability of VM-based MSFC systems. At present, many studies have used analytical models to evaluate the availability and reliability of SFC systems.

i). Non-state-space model

Wang *et al.* [19] studied SFC's availability based on RBD, and calculated the steady-state availability of the system with series/parallel configuration. It is noticed that these methods cannot analyze the time dependence between individual components.

ii). State-space model

In [20], Mauro *et al.* proposed a stochastic reward net (SRN) for SFC availability evaluation. In [21], Li *et al.* proposed a Petri net approach to capture the behaviors and dependencies of all network elements in 5G and beyond telecommunication networks for resilience evaluation and optimization. Simone *et al.* [22] carried out a latency-driven availability assessment for multi-tenant service chains through CTMC. They developed a modified version of the multidimensional universal generating function technique to compute the availability. The above studies only considered the behaviors in terms of failure/repair events and didn't consider the impact caused by software aging. In contrast, our model captures component behaviors in VM-based MSFC from suffering software aging to recovering.

In [23], Tola *et al.* proposed an availability model based on stochastic activity networks (SANs), for cloud-native management and orchestration implementations. The model is further extended in [24] to incorporate software rejuvenation techniques against software aging. The researches of [23][24] considered the impact of software aging, but CTMC-based methods require that all time intervals must follow exponential distribution. However, in many cases, there exist one or more time intervals that do not follow exponential distribution.

In [25] and [26], the authors analyzed the dependability of the serial and serial-parallel hybrid SFC services. They later considered the impact of the rejuvenation trigger intervals in [27]. They further developed SMP models in [28] and [29] for capturing the behaviors of multiple SFs and operating systems (OSes) in serial SFC system and serial-parallel hybrid SFC system, respectively. In addition, they analyzed the impact of backup component behaviors on SFC dependability in [30]. The authors of [25]-[30] considered the impact of software aging and assumed that the failure and recovery time followed the general distributions. But the models proposed in these studies were not suitable to capture the interaction behaviors of SFs, VMs, and VMMs in VM-based MSFC.

iii). Hierarchical model

In [31], Pathirana *et al.* proposed a two-level model to calculate the availability of 5G-MEC systems. At the higher level, they used a FT model to capture the interdependencies among the various 5G and MEC elements. At the lower layer, they used SANs to capture multiple failure modes for each element in the 5G-MEC system. In the hierarchical models, FT and RBD are usually used as the higher level model, and CTMC model is usually used as the lower level model. The dynamic model for each component is usually considered to be independent. However, there may be interactions between the behaviors of individual components, resulting in invalidation of the higher level model. For example, when considering the live migration technique, the state of the component will be interdependent with other components.

TABLE I gives the comparison of the above related studies and our work with the following three major differences:

- The models proposed by studies [13], [16]-[18], [20]-[24] and [31] assumed that the occurrence time of all events followed exponential distribution. The models established in studies [13]-[31] were not suitable for evaluating the dependability of VM-based MSFC. The models in studies [13]-[18] involved a single SF and cannot be directly used to capture the behaviors of SFC systems with multiple SFs. Different from these models, our model can analyze the time-dependent interactions between the abnormal behaviors and the recovery behaviors of components in VM-based MSFC under the condition that the occurrence times of failure events and recovery events follow any type of distribution.
- Studies [14]-[16], [19]-[21], [23], [24] and [31] did not analyze the steady-state availability and reliability together. Different from these studies, we derive the formulas for calculating the steady-state availability and reliability of VM-based MSFC in the system with an arbitrary number of SFs, VMs, and VMMs, to analyze the effectiveness of proactive rejuvenation techniques from multiple dimensions.
- The authors in [13]-[24] and [31] did not conduct simulation experiments and there is no sensitivity analysis in [13]-[15], [19]-[25] and [31]. Different from these studies, we first validate our proposed model and metric formulas via simulation. In addition, through the sensitivity analysis, we find the bottlenecks restricting the effectiveness of proactive rejuvenation techniques, laying a foundation for optimizing the availability and reliability of VM-based MSFC.

# 3 VM-Based MSFC Dependability Model Description

This section first introduces a VM-based MSFC system considered in this paper. Then the proposed SMP is described.

## 3.1 System Description

A VM-based MSFC system consists of a higher-level controller, several domain controllers, several Primary Hosts and several Backup Hosts. The domain controller is deployed to manage the resources of the computation and bandwidth of the domain. The higher-level controller manages resources among domains and domain controllers. Primary Host contains an active VMM (AVMM), which hosts multiple active VMs (AVMs) and backup VMs (BVMs). Active SFs (ASFs) are executed in AVMs and backup SFs (BSFs) are executed in BVMs for supporting failover technique. Backup Hosts are used to support live VM migration technique. SF, VM and VMM are all software and all of them can suffer from software aging and failure due to software aging.

TABLE I
COMPARISON OF THE EXISTING STUDIES DISCUSSED IN SECTION 2

| Ref. | SFC | SF | VM | VMM | Distribution | | Evaluation Metric | | Solution Technique | | Sensitivity Analysis | Model | Time-Dependence among SFs, VMs and VMMs | Aging Behavior |
|---|---|---|---|---|---|---|---|---|---|---|---|---|---|---|
| | | | | | Exponential | General | Availability | Reliability | Simulation | Modeling | | | | |
| **Zhu 2020 [13]** | × | × | √ | √ | √ | × | √ | √ | × | √ | × | CTMC | × | √ |
| **Martyna 2021 [14]** | × | × | √ | × | √ | √ | √ | × | × | √ | × | SMP | × | × |
| **Bai 2022 [15]** | × | √ | √ | √ | √ | √ | √ | × | × | √ | × | MRGP | × | √ |
| **Zhang 2022 [16]** | × | √ | √ | × | √ | × | √ | × | × | √ | √ | CTMC | × | × |
| **Torquato 2022 [17]** | × | × | √ | × | √ | × | √ | √ | × | √ | √ | CTMC | × | √ |
| **Clemente 2022 [18]** | × | √ | √ | × | √ | × | √ | √ | × | √ | √ | CTMC | × | × |
| **Wang 2021 [19]** | √ | √ | × | × | × | × | √ | × | × | √ | × | CTMC | × | × |
| **Mauro 2020 [20]** | √ | √ | × | √ | √ | × | √ | × | × | √ | × | CTMC | × | × |
| **Li 2022 [21]** | √ | √ | × | × | √ | × | √ | × | × | √ | × | CTMC | × | × |
| **Simone 2023 [22]** | √ | √ | √ | × | √ | × | √ | √ | × | √ | × | CTMC | × | × |
| **Tola 2020 [23], 2021 [24]** | √ | √ | √ | × | √ | × | √ | × | × | √ | × | CTMC | × | √ |
| **Bai 2021 [25]** | √ | √ | × | × | √ | √ | √ | √ | √ | √ | × | SMP | × | √ |
| **Bai 2022 [26][27]** | √ | √ | × | × | √ | √ | √ | √ | √ | √ | √ | SMP | × | √ |
| **Bai 2023 [28]-[30]** | √ | √ | × | √ | √ | √ | √ | √ | √ | √ | √ | SMP | × | √ |
| **Pathirana 2023 [31]** | √ | √ | √ | √ | √ | × | √ | × | × | √ | × | CTMC | × | × |
| **Our work** | √ | √ | √ | √ | √ | √ | √ | √ | √ | √ | √ | SMP | √ | √ |

TABLE II
REJUVENATION TYPES

| Object | Rejuvenation Type | Number of Aging Components | Rejuvenation Technique | Number of Components to be Recovered |
|---|---|---|---|---|
| **SF** | Single | One ASF | SF Failover | One ASF |
| | Host | Multiple ASFs on a Primary Host | Restart | All SFs executed on Primary Host running the aging ASFs |
| | System | Multiple ASFs on different Primary Hosts | Restart | All SFs |
| **VM** | Single | One AVM | VM Failover | One AVM |
| | Portion | One AVM and the ASF executed on this AVM | Restart | One AVM and all SFs executed on this AVM |
| | Host | Multiple AVMs or one ASF and AVM executing other ASFs on the same Primary Host | Restart | All ASFs and AVMs executed on this Primary Host |
| | System | Multiple AVMs running on different Primary Host | Restart | All SFs and VMs |
| **VMM** | Single | One AVMM | VM Migration | One AVMM |
| | Host | One AVMM and one ASF/AVM executed on this AVMM | Restart/ Reboot | AVMM and all SFs, and VMs executed on this AVMM |
| | System | Multiple AVMMs | Restart/Reboot | All SFs, VMs and VMMs |

Based on the number of aging ASFs, we consider three types of SF rejuvenation, as shown in TABLE II. In the SF-Single type, if an ASF is detected to suffer from software aging, the SF failover technique is triggered. In the SF-Host type, if multiple ASFs running on a Primary Host are detected to suffer from software aging, all SFs on this host are restarted. In the SF-System type, if multiple ASFs running on different Primary Hosts are detected to suffer from software aging, all SFs in the system are restarted.

Based on the number of aging AVMs, we consider four types of VM rejuvenation, as shown in TABLE II.

- In the VM-Single type, if an AVM is detected to suffer from software aging, the VM failover technique is triggered.
- In the VM-Portion type, if an AVM and the ASF running in it are detected to suffer from software aging, this AVM is restarted.
- In the VM-Host type, if multiple AVMs running on the same Primary Host are detected to suffer from software aging or one ASF and AVM executing other ASFs on the same Primary Host are detected to suffer from software aging, all SFs and VMs on this Primary host are restarted.
- In the VM-System type, after an AVM running on one Primary Host suffers from software aging, if AVMs or ASFs running on other Primary Hosts are detected to suffer from software aging, all SFs and VMs in the system are restarted.

Based on the number of aging AVMMs, we consider three types of VMM rejuvenation, as shown in TABLE II.

- In the VMM-Single type, if an AVMM is detected to suffer from software aging, live VM migration technique are employed immediately.
- In the VMM-Host type, after an AVMM running on one Primary Host suffers from software aging, if ASFs or AVMs running on this host are detected to suffer from software aging, all SFs, VMs and VMM on this host are restarted/rebooted.
- In the VMM-System type, if multiple AVMMs are detected to suffer from software aging, all SFs, VMs and VMMs in the system are restarted/rebooted.

If one of the ASFs, AVMs, and AVMMs fails, the VM-based MSFC is unavailable. After the failed component is fixed, all SFs, VMs, and VMMs in the VM-based MSFC system will be restarted/rebooted. We assume that the holding times of aging events are exponentially distributed, and the holding times of other events follow general distribution.

### 3.2 SMP Model

We define ($n$+2$m$)-tuple index $(i_{\mathrm{sf}(1)},\ldots,i_{\mathrm{sf}(m)},j_{\mathrm{vm}(1)},j_{\mathrm{vm}(2)},j_{\mathrm{vm}(3)},\ldots,$ $j_{\mathrm{vm}(m)},k_{\mathrm{vmm}(1)},k_{\mathrm{vmm}(2)},k_{\mathrm{vmm}(3)},\ldots,k_{\mathrm{vmm}(n)})$ to denote system state. Here, $n, m \in \mathrm{N}^*$. $i_{\mathrm{sf}m}$, $j_{\mathrm{vm}m}$ and $k_{\mathrm{vmm}n}$ denote the states of the $m^{\mathrm{th}}$ ASF, the $m^{\mathrm{th}}$ AVM and the $n^{\mathrm{th}}$ AVMM, respectively. There are four states: Perfect, Unstable, Recovery and Failed, denoted by P, U, R and F, respectively. The following details each state:

- State P (Perfect): In this state, the system execution is normal. The system works well for users. Recovery operation can make the system back to this state.
- State U (Unstable): In this state, the processing efficiency decreases and the possibility of system failure increases due to software aging. But the system is still available to users. If no recovery operation is adopted, the system will fail.
- State R (Recovery): In this state, recovery operation is used.
- State F (Failed): In this state, VM-based MSFC is unavailable.

The underlying system behaviors can be described by a SMP $\{Z(t) \mid t \geq 0\}$ . $\mathrm{S}=\{S_0,S_1,\ldots \quad \ldots \quad {}_n\}$ is defined to denote the system state space. Each $S_i$ represents a system state. The state sequence of the stochastic process at the transition time points forms an embedded discrete-time Markov chain (EDTMC). Fig. 3 shows an example of the SMP model for the VM-based MSFC system with four ASFs, four AVMs and two AVMMs. Note that the proposed model in this paper can be used to analyze the VM-based MSFC system with $m$ ASFs, $m$ AVMs and $n$ AVMMs. TABLE III shows the definition of variables used in the model.

In our SMP model, the conditions for each state transition is described in Section 3.1. The VM-based MSFC starts from state $S_0$=($\mathrm{P}_{\mathrm{sf}(1)},\ldots,\mathrm{P}_{\mathrm{sf}(m)},\mathrm{P}_{\mathrm{vm}(1)},\ldots,\mathrm{P}_{\mathrm{vm}(m)},\mathrm{P}_{\mathrm{vmm}(1)},\ldots,\mathrm{P}_{\mathrm{vmm}(n)}$). After a period of execution, the components can suffer from software aging. If the first ASF is detected to suffer from software aging, the system state moves from state $S_0$ to state $S_{4+i} = (\mathrm{P}_{\mathrm{sf}(1)},\ldots,\mathrm{U}_{\mathrm{sf}(i)},\ldots,\mathrm{P}_{\mathrm{sf}(m)},\mathrm{P}_{\mathrm{vm}(1)},\ldots,\mathrm{P}_{\mathrm{vm}(m)},\mathrm{P}_{\mathrm{vmm}(1)},\ldots)$ . In this case, the rejuvenation technique of the SF-Single type is adopted, that is, the BSF takes over the processing of requests. The rejuvenation operations used in the state-transition process are shown in TABLE IV. In addition, if any component is detected to fail, the VM-based MSFC enters Failed state $S_1$=($\mathrm{F}_{\mathrm{sf}(1)},\ldots,\mathrm{F}_{\mathrm{sf}(m)},\mathrm{F}_{\mathrm{vm}(1)},\mathrm{F}_{\mathrm{vm}(2)},\ldots,\mathrm{F}_{\mathrm{vm}(m)},\mathrm{F}_{\mathrm{vmm}(1)},\mathrm{F}_{\mathrm{vmm}(2)},\ldots,\mathrm{F}_{\mathrm{vmm}(n)}$). All components will be restarted/rebooted after the failed component is fixed, then the system enters the Perfect state $S_0$.

## 4 MSFC Dependability Analysis

As dependability measures, we consider the steady-state availability and reliability of VM-based MSFC.

### 4.1 Steady-State Availability Analysis

This section describes the process of calculating the steady-state availability of VM-based MSFC.

First, we construct the kernel matrix $\mathbf{K}(t)$. We put the kernel matrix $\mathbf{K}(t)$ and the non-null elements of it in Section A of the supplementary file due to the limitation of space. By solving the one-step transition probability matrix (TPM) $\mathbf{P}=\left[P_{SiSj}\right]=\lim_{t\to\infty}\mathbf{K}(t)$ , we can characterize the EDTMC of the SMP. The one-step TPM $\mathbf{P}$ are given in Section A of the supplementary file. Secondly, in order to obtain the steady-state probability vector $\mathbf{V}$ of the EDTMC, we solve the linear equations, as shown in Equation (1).

$$\mathbf{V} = \mathbf{VP} \text{ subject to } \mathbf{V}e^{\mathrm{T}} = 1 \tag{1}$$

The equations of calculating the steady-state probability $V_{S_i}$ of the EDTMC in system state $P_{S_i}$ are given in Section A of the supplementary file. Next, we need to calculate the average sojourn time $h_{S_i}$ in system state $S_i$, as shown in Equation (2). The calculation formula of the mean sojourn time in each state is shown in Section A of the supplementary file.

$$h_{S_i} = \int_0^{\infty}(1 - G_{S_i}(t))dt \tag{2}$$

The steady-state probability $\pi_{S_i}$ in system state $S_i$ is calculated according to Equation (3):

$$\pi_{S_i} = \frac{v_{S_i} h_{S_i}}{\sum_{S_j} v_{S_j} h_{S_j}} \tag{3}$$

where $v_{S_i}$ and $h_{S_i}$ can be obtain by Equations (1-2). The availability $A_s$ of VM-based MSFC consisting of $m$ ASFs executed in $n$ AVMM can be obtained from the sum of the steady-state probabilities of the states $S_0$, $S_{4+i}$, $S_{4+m+i}$ and $S_{4+2m+i}$, as shown in Equation (4).

$$A = \pi_0 + \sum_{i=1}^{m}\pi_{4+i} + \sum_{i=1}^{m}\pi_{4+m+i} + \sum_{i=1}^{n}\pi_{4+2m+i} \tag{4}$$

### 4.2 Reliability Analysis

This section deduces the formula of calculating the reliability of VM-based MSFC, evaluated based on the mean time to VM-based MSFC failure (MTTF). It is necessary to consider a system where there is no recovery operation when MSFC is unavailable. First, we construct a kernel matrix $\mathbf{K}'(t)$ for the SMP model with absorbing states. $\mathbf{K}'(t)$ can be represented as Fig. B-1 in the supplementary file. The non-null elements of the $\mathbf{K}'(t)$ under this model are calculated by Equations (5)-(10).

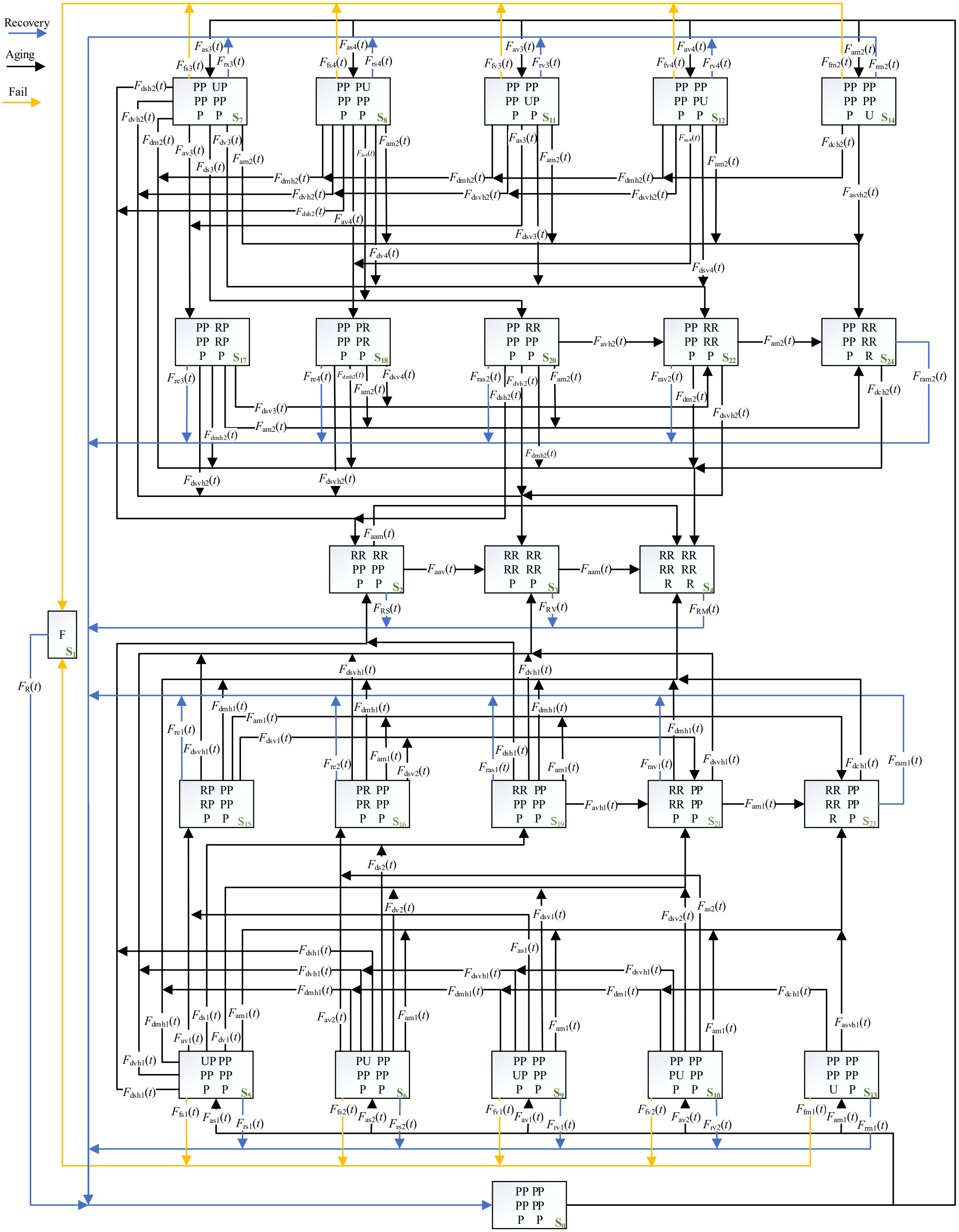

**Fig. 3** SMP model for describing the behaviors of VM-based MSFC system with four ASFs, four AVMs and two AVMMs

TABLE III
DEFINITIONS OF VARIABLES USED IN THE MODEL

| Symbol | Definition |
|---|---|
| $T_{asi}$ | Random variable with cumulative distribution function (CDF) $F_{asi}(t)$ denoting the holding time of the ith ASF from P state to U state. |
| $T_{avi}$ | Random variable with CDF $F_{avi}(t)$ denoting the holding time of the $i^{th}$ AVM from P state to U state. |
| $T_{ami}$ | Random variable with CDF $F_{ami}(t)$ denoting the holding time of the $i^{th}$ AVMM from P state to U state. |
| $T_{rsi}$ | Random variable with CDF $F_{rsi}(t)$ denoting the holding time of the $i^{th}$ ASF from U state to P state by SF failover technique. (SF failover time) |
| $T_{rvi}$ | Random variable with CDF $F_{rvi}(t)$ denoting the holding time of the $i^{th}$ AVM from U state to P state by VM failover technique. (VM failover time) |
| $T_{rmi}$ | Random variable with CDF $F_{rmi}(t)$ denoting the holding time of the $i^{th}$ AVMM from U state to P state by VM migration technique. (VM migration time) |
| $T_{rei}$ | Random variable with CDF $F_{rei}(t)$ denoting the holding time of restarting the $i^{th}$ AVM. |
| $T_{rasi}$ | Random variable with CDF $F_{rasi}(t)$ denoting the holding time of restarting all ASFs on the $i^{th}$ Primary Host. |
| $T_{ravi}$ | Random variable with CDF $F_{ravi}(t)$ denoting the holding time of restarting all ASFs and AVMs in the ith Primary Host. |
| $T_{rami}$ | Random variable with CDF $F_{rami}(t)$ denoting the holding time of rebooting the $i^{th}$ Primary Host. |
| $T_{RS}$ | Random variable with CDF $F_{RS}(t)$ denoting the holding time of restarting all ASFs in the system. |
| $T_{RV}$ | Random variable with CDF $F_{RV}(t)$ denoting the holding time of restarting all ASFs and AVMs in the system. |
| $T_{RM}$ | Random variable with CDF $F_{RM}(t)$ denoting the holding time of rebooting all AVMMs. |
| $T_{R}$ | Random variable with CDF $F_{R}(t)$ denoting the holding time of repairing and rebooting VM-based MSFC system. |
| $T_{fsi}$ | Random variable with CDF $F_{fsi}(t)$ denoting the holding time of the $i^{th}$ ASF from U state to F state. |
| $T_{fvi}$ | Random variable with CDF $F_{fvi}(t)$ denoting the holding time of the $i^{th}$ AVM from U state to F state. |
| $T_{fmi}$ | Random variable with CDF $F_{fmi}(t)$ denoting the holding time of the $i^{th}$ AVMM from U state to F state. |
| $T_{dshi}$ | Random variable with CDF $F_{dshi}(t)$ denoting the minimum holding time of ASFs on other Primary Host from P state to U state after one ASF or more than one ASF on the $i^{th}$ Primary Host suffering from software aging. |
| $T_{dvhi}$ | Random variable with CDF $F_{dvhi}(t)$ denoting the minimum holding time of AVMs on other Primary Host from P state to U state after one ASF or more than one ASF on the $i^{th}$ Primary Host suffering from software aging. |
| $T_{dmhi}$ | Random variable with CDF $F_{dmhi}(t)$ denoting the minimum holding time of AVMMs on other Primary Host from P state to U state after at least one ASF/AVM on the $i^{th}$ Primary Host having software aging or after an ASF and an AVMs on the $i^{th}$ Primary Host having software aging. |
| $T_{dsvhi}$ | Random variable with CDF $F_{dsvhi}(t)$ denoting the minimum holding time of ASFs and AVMs on other Primary Host from P state to U state after one or more than one AVM in the $i^{th}$ Primary Host suffering from software aging or after one of ASFs and one of AVMs on the $i^{th}$ Primary Host suffering from software aging. |
| $T_{dchi}$ | Random variable with CDF $F_{dchi}(t)$ denoting the minimum holding time of ASFs, AVMs and AVMMs on other Primary Host from P state to U state after the AVMM suffering from software aging on the $i^{th}$ Primary Host. |
| $T_{dsi}$ | Random variable with CDF $F_{dsi}(t)$ denoting the minimum holding time of other ASFs from P state to U state on the Primary Host where the ith ASF is running after the ith ASF suffering from software aging. |
| $T_{dvi}$ | Random variable with CDF $F_{dvi}(t)$ denoting the minimum holding time of AVMs, except for the $i^{th}$ AVM, from P state to U state on the Primary Host where the ith ASF is running after the ith ASF suffering from software aging. |
| $T_{dsvi}$ | Random variable with CDF $F_{dsvi}(t)$ denoting the minimum holding time of ASFs and AVMs, except for the $i^{th}$ ASF and the $i^{th}$ AVM, from P state to U state on the Primary Host where the $i^{th}$ ASF and the $i^{th}$ AVM is running after the $i^{th}$ AVM suffering from software aging or after both the ith ASF and the ith AVM suffering from software aging. |
| $T_{asvhi}$ | Random variable with CDF $F_{asvhi}(t)$ denoting the minimum holding time of ASFs and AVMs on the $i^{th}$ Primary Host from P state to U state after the $i^{th}$ AVMM suffering from software aging. |
| $T_{avhi}$ | Random variable with CDF $F_{avhi}(t)$ denoting the minimum holding time of AVMs on the $i^{th}$ Primary Host from P state to U state after more than one ASF suffering from software aging on the $i^{th}$ Primary Host. |
| $T_{aav}$ | Random variable with CDF $F_{aav}(t)$ denoting the minimum holding time of AVMs in the system from P state to U state after multiple ASFs suffering from software aging on multiple Primary Hosts. |
| $T_{aam}$ | Random variable with CDF $F_{aam}(t)$ denoting the minimum holding time of AVMMs from P state to U state after multiple ASFs/AVMs being in software aging on multiple Primary Hosts or after multiple ASFs and AVMs suffering from software aging on multiple Primary Hosts. |

TABLE IV
REJUVENATION TECHNIQUES ADOPTED FOR STATE TRANSITIONS

| State Transition | Rejuvenation Type |
|---|---|
| From $S_0(P_{sf(1)},...,P_{sf(m)},P_{vm(1)},...,P_{vm(m)},P_{vmm(1)},...,P_{vmm(n)})$ to $S_{4+i}(P_{sf(1)},...,U_{sf(i)},...,P_{sf(m)},P_{vm(1)},...,P_{vm(m)},P_{vmm(1)},...,P_{vmm(n)})$ | SF-Single |
| From $S_{4+i}(P_{sf(1)},...,U_{sf(i)},...,P_{sf(m)},P_{vm(1)},...,P_{vm(m)},P_{vmm(1)},...)$ to $S_{4+3m+n+j}(P_{sf(1)},...,R_{sf(host_j)},...,P_{sf(m)},P_{vm(1)},...,P_{vm(m)},P_{vmm(1)},...)$ | SF-Host |
| From $S_{4+i}(P_{sf(1)},...,U_{sf(i)},...,P_{sf(m)},P_{vm(1)},...,P_{vm(m)},P_{vmm(1)},...,P_{vmm(n)})$ to | SF-System |
| From $S_0(P_{sf(1)},...,P_{sf(m)},P_{vm(1)},...,P_{vm(m)},P_{vmm(1)},...,P_{vmm(n)})$ to $S_{4+m+i}(P_{sf(1)},...,P_{sf(m)},P_{vm(1)},...,U_{vm(i)},...,P_{vm(m)},P_{vmm(1)},...,P_{vmm(n)})$ | VM-Single |
| From $S_{4+m+i}(P_{sf(1)},...,P_{sf(m)},P_{vm(1)},...,U_{vm(i)},...,P_{vm(m)},P_{vmm(1)},...)$ to $S_{4+2m+n+i}(P_{sf(1)},...,U_{sf(i)},...,P_{sf(m)},P_{vm(1)},...,U_{vm(i)},...,P_{vm(m)},P_{vmm(1)},...)$ | VM-Portion |
| From $S_{4+m+i}(P_{sf(1)},...,P_{vm(1)},...,U_{vm(i)},...,P_{vm(m)},P_{vmm(1)},...)$ to $S_{4+3m+2n+j}(P_{sf(1)},...,R_{sf(host_j)},...,P_{sf(m)},P_{vm(1)},...,R_{vm(host_j)},...,P_{vm(m)},P_{vmm(1)},...)$ | VM-Host |
| From $S_{4+m+i}(P_{sf(1)},...,P_{sf(m)},P_{vm(1)},...,U_{vm(i)},...,P_{vm(m)},P_{vmm(1)},...,P_{vmm(n)})$ to $S_3(R_{sf(1)},...,R_{sf(m)},R_{vm(1)},...,R_{vm(m)},P_{vmm(1)},...,P_{vmm(n)})$ | VM-System |
| From $S_0(P_{sf(1)},...,P_{sf(m)},P_{vm(1)},...,P_{vm(m)},P_{vmm(1)},...,P_{vmm(n)})$ to $S_{4+2m+j}(P_{sf(1)},...,P_{sf(m)},P_{vm(1)},...,P_{vm(m)},P_{vmm(1)},...,U_{vmm(j)},...,P_{vmm(n)})$ | VMM-Single |
| From $S_{4+2m+j}(P_{sf(1)},...P_{vmm(1)},...,U_{vmm(j)},...,P_{vmm(n)})$ to $S_{4+3m+3n+j}(P_{sf(1)},...,R_{sf(host_j)},...,P_{sf(m)},P_{vm(1)},...,R_{vm(host_j)},...,P_{vmm(1)},...,R_{vmm(j)},...,P_{vmm(n)})$ | VMM-Host |
| From $S_{4+2m+j}(P_{sf(1)},...,P_{sf(m)},P_{vm(1)},...,P_{vm(m)},P_{vmm(1)},...,U_{vmm(j)},...,P_{vmm(n)})$ to $S_4(R_{sf(1)},...,R_{sf(m)},R_{vm(1)},...,R_{vm(m)},R_{vmm(1)},...,R_{vmm(n)})$ | VMM-System |

* The Subscript $sf(host_j)$ / $vm(host_j)$ denotes the number of the SF/VM on the $j^{th}$ host.

* $1 \le i \le m, 1 \le j \le n$

$$k'_{S_0^{'}S_i^{'}}(t) = \int_0^t \prod_{w=1}^{m}(1-F_{\mathrm{av}w}(t))\prod_{y=1}^{n}(1-F_{\mathrm{am}y}(t))\prod_{x\in A_i^{'}}(1-F_{\mathrm{as}x}(t))dF_{\mathrm{as}i}(t) \tag{5}$$

$$k'_{S_0^{'}S_{m+i}^{'}}(t) = \int_0^t \prod_{w=1}^{m}(1-F_{\mathrm{as}w}(t))\prod_{y=1}^{n}(1-F_{\mathrm{am}y}(t))\prod_{x\in A_i^{'}}(1-F_{\mathrm{av}x}(t))dF_{\mathrm{av}i}(t) \tag{6}$$

$$k'_{S_0^{'}S_{2m+j}^{'}}(t) = \int_0^t \prod_{w=1}^{n}(1-F_{\mathrm{as}w}(t))\prod_{x=1}^{m}(1-F_{\mathrm{av}x}(t))\prod_{y\in B_j^{'}}(1-F_{\mathrm{am}y}(t))dF_{\mathrm{am}j}(t) \tag{7}$$

$$k'_{S_i^{'}S_0^{'}}(t) = \int_0^t (1-F_{\mathrm{fs}i}(t))(1-F_{\mathrm{dsh}u}(t))(1-F_{\mathrm{dvh}u}(t))(1-F_{\mathrm{dmh}u}(t))(1-F_{\mathrm{av}i}(t))(1-F_{\mathrm{ds}i}(t))(1-F_{\mathrm{dv}i}(t))(1-F_{\mathrm{am}u}(t))dF_{\mathrm{rs}i}(t) \tag{8}$$

$$k'_{S_{m+i}^{'}S_0^{'}}(t) = \int_0^t (1-F_{\mathrm{fv}i}(t))(1-F_{\mathrm{dmh}u}(t))(1-F_{\mathrm{dsvh}u}(t))(1-F_{\mathrm{dsv}i}(t))(1-F_{\mathrm{as}i}(t))(1-F_{\mathrm{am}u}(t))dF_{\mathrm{rv}i}(t) \tag{9}$$

$$k'_{S_{2m+j}^{'}S_0^{'}}(t) = \int_0^t (1-F_{\mathrm{fm}j}(t))(1-F_{\mathrm{dch}j}(t))(1-F_{\mathrm{asvh}j}(t))dF_{\mathrm{rm}j}(t) \tag{10}$$

where $1\le i\le m$ , $1\le j\le n$ , $A_i^{'}=\{x\,|\,1\le x\le m, x\ne i\}$ , $B_j^{'}=\{y\,|\,1\le y\le n, y\ne j\}$ and $u$ means that the $u^{\mathrm{th}}$ Primary Host that executing the $i^{\mathrm{th}}$ ASF/AVM suffering from software aging. Next, according to the method mentioned in the previous section, we can obtain a one-step TPM $\mathbf{P}'$ for describing the EDTMC in the SMP with absorption states.

The expected number of visits $V_{S_i^{'}}^{*}$ to system state $S_i^{'}$ until absorption is calculated by applying Equation (11).

$$V_{S_i^{'}}^{*} = \alpha_{S_i^{'}} + \sum_{j=0}^{2m+n} V_{S_j^{'}}^{*} p'_{S_j^{'}S_i^{'}} \tag{11}$$

where $\alpha_{S_i^{'}}$ is the initial probability in system state $S_i^{'}$. $V_{S_0^{'}}^{*}$ and $V_{S_i^{'}}^{*}$ $(1\le i\le 2m+n)$ are given by Equations (12) and (13), respectively.

$$V_{S_0^{'}}^{'} = (-1)/(\sum_{i=1}^{2m+n} p'_{S_0^{'}S_i^{'}}\, p'_{S_i^{'}S_0^{'}} - 1) \tag{12}$$

$$V_{S_i^{'}}^{'} = (-p'_{S_0^{'}S_i^{'}})/(\sum_{i=1}^{2m+n} p'_{S_0^{'}S_i^{'}}\, p'_{S_i^{'}S_0^{'}} - 1) \tag{13}$$

The mean sojourn time $h_{S_i^{'}}^{'}$ in system state $S_i^{'}$ are calculated in Equations (14)-(17).

$$h'_{S_0^{'}} = \int_0^{\infty} \prod_{w=1}^{\mathrm{m}}(1-F_{\mathrm{as}w}(t))\prod_{x=1}^{\mathrm{m}}(1-F_{\mathrm{av}x}(t))\prod_{y=1}^{\mathrm{n}}(1-F_{\mathrm{am}y}(t))dt \tag{14}$$

$$h'_{S_i^{'}} = \int_0^{\infty} (1-F_{\mathrm{fs}i}(t))(1-F_{\mathrm{rs}i}(t))(1-F_{\mathrm{dsh}u}(t))(1-F_{\mathrm{dv}i}(t))(1-F_{\mathrm{am}u}(t))(1-F_{\mathrm{dvh}u}(t))(1-F_{\mathrm{dmh}u}(t))(1-F_{\mathrm{av}i}(t))(1-F_{\mathrm{ds}i}(t))dt \tag{15}$$

$$h'_{S_{m+i}^{'}} = \int_0^{\infty} (1-F_{\mathrm{fv}i}(t))(1-F_{\mathrm{rv}i}(t))(1-F_{\mathrm{dmh}u}(t))(1-F_{\mathrm{dsvh}u}(t))(1-F_{\mathrm{dsv}i}(t))(1-F_{\mathrm{as}i}(t))(1-F_{\mathrm{am}u}(t))dt \tag{16}$$

$$h'_{S_{2m+j}^{'}} = \int_0^{\infty} (1-F_{\mathrm{fm}j}(t))(1-F_{\mathrm{rm}j}(t))(1-F_{\mathrm{dch}j}(t))(1-F_{\mathrm{asvh}j}(t))dt \tag{17}$$

where $1\le i\le m$ and $1\le j\le n$ .

TABLE V

TYPES OF CDFS USED IN THE EXPERIMENTS

| Symbol | CDF | Default Value | Symbol | CDF | Default Value |
|---|---|---|---|---|---|
| $T_{\mathrm{as}i}$ | $F_{\mathrm{as}i}(t)=EXP(\alpha_{\mathrm{s}i}), i\in[1,m]$ | 9-11 days | $T_{\mathrm{fs}i}$ | $F_{\mathrm{fs}i}(t)=HYPO(\lambda_{\mathrm{s}i1},\lambda_{\mathrm{s}i2}), i\in[1,m]$ | 6-8 days |
| $T_{\mathrm{av}i}$ | $F_{\mathrm{av}i}(t)=EXP(\alpha_{\mathrm{v}i}), i\in[1,m]$ | 15-17 days | $T_{\mathrm{fv}i}$ | $F_{\mathrm{fv}i}(t)=HYPO(\lambda_{\mathrm{v}i1},\lambda_{\mathrm{v}i2}), i\in[1,m]$ | 15-17 days |
| $T_{\mathrm{am}i}$ | $F_{\mathrm{am}i}(t)=EXP(\alpha_{\mathrm{m}i}), i\in[1,n]$ | 20-25 days | $T_{\mathrm{fm}i}$ | $F_{\mathrm{fm}i}(t)=HYPO(\lambda_{\mathrm{m}i1},\lambda_{\mathrm{m}i2}), i\in[1,n]$ | 30-32 days |
| $T_{\mathrm{rs}i}$ | $F_{\mathrm{rs}i}(t)=EXP(\beta_{\mathrm{s}i}), i\in[1,m]$ | 8-11 seconds | $T_{\mathrm{rv}i}$ | $F_{\mathrm{rv}i}(t)=EXP(\beta_{\mathrm{v}i}), i\in[1,m]$ | 15-17 seconds |
| $T_{\mathrm{rm}i}$ | $F_{\mathrm{rm}i}(t)=EXP(\beta_{\mathrm{m}i}), i\in[1,n]$ | 30-31 seconds | $T_{\mathrm{ras}i}$ | $F_{\mathrm{ras}i}(t)=EXP(\gamma_{\mathrm{s}i}), i\in[1,n]$ | 30-31 seconds |
| $T_{\mathrm{re}i}$ | $F_{\mathrm{re}i}(t)=EXP(\gamma_{\mathrm{e}i}), i\in[1,m]$ | 11-13 seconds | $T_{\mathrm{rav}i}$ | $F_{\mathrm{rav}i}(t)=EXP(\gamma_{\mathrm{v}i}), i\in[1,n]$ | 45-46 seconds |
| $T_{\mathrm{R}}$ | $F_{\mathrm{R}}(t)=EXP(\varepsilon)$ | 0.8-1 hours | $T_{\mathrm{RS}}$ | $F_{\mathrm{RS}}(t)=EXP(\delta_{\mathrm{s}})$ | 1-1.1 minutes |
| $T_{\mathrm{RV}}$ | $F_{\mathrm{RV}}(t)=EXP(\delta_{\mathrm{v}})$ | 2-2.1 minutes | $T_{\mathrm{RM}}$ | $F_{\mathrm{RM}}(t)=EXP(\delta_{\mathrm{m}})$ | 3-3.1 minutes |
| $T_{\mathrm{ram}i}$ | $F_{\mathrm{ram}i}(t)=EXP(\gamma_{\mathrm{m}i}), i\in[1,n]$ | 1-1.1 minutes | $T_{\mathrm{aav}}$ | $F_{\mathrm{aav}}(t)=EXP(\sum_{k\in E}\alpha_{\mathrm{v}k})$ | -- |
| $T_{\mathrm{dsvh}i}$ | $F_{\mathrm{dsvh}i}(t)=EXP(\sum_{k\in A_i}(\alpha_{\mathrm{s}k}+\alpha_{\mathrm{v}k})), i\in[1,n]$ | -- | $T_{\mathrm{dsv}i}$ | $F_{\mathrm{dsv}i}(t)=EXP(\sum_{k\in C_i}(\alpha_{\mathrm{s}k}+\alpha_{\mathrm{v}k})), i\in[1,m]$ | -- |
| $T_{\mathrm{dv}i}$ | $F_{\mathrm{dv}i}(t)=EXP(\sum_{k\in C_i}\alpha_{\mathrm{v}k}), i\in[1,m]$ | -- | $T_{\mathrm{asvh}i}$ | $F_{\mathrm{asvh}i}(t)=EXP(\sum_{k\in D_i}(\alpha_{\mathrm{s}k}+\alpha_{\mathrm{v}k})), i\in[1,n]$ | -- |
| $T_{\mathrm{ds}i}$ | $F_{\mathrm{ds}i}(t)=EXP(\sum_{k\in C_i}\alpha_{\mathrm{s}k}), i\in[1,m]$ | -- | $T_{\mathrm{avh}i}$ | $F_{\mathrm{avh}i}(t)=EXP(\sum_{k\in D_i}\alpha_{\mathrm{v}k}), i\in[1,n]$ | -- |
| $T_{\mathrm{dsh}i}$ | $F_{\mathrm{dsh}i}(t)=EXP(\sum_{k\in A_i}\alpha_{\mathrm{s}k}), i\in[1,n]$ | -- | $T_{\mathrm{dch}i}$ | $F_{\mathrm{dch}i}(t)=EXP(\sum_{k\in A_i}(\alpha_{\mathrm{s}_k}+\alpha_{\mathrm{v}_k})+\sum_{k\in B_i}\alpha_{\mathrm{m}_k}), i\in[1,n]$ | -- |
| $T_{\mathrm{dvh}i}$ | $F_{\mathrm{dvh}i}(t)=EXP(\sum_{k\in A_i}\alpha_{\mathrm{v}k}), i\in[1,n]$ | -- | $T_{\mathrm{dmh}i}$ | $F_{\mathrm{dmh}i}(t)=EXP(\sum_{k\in B_i}\alpha_{\mathrm{m}k}), i\in[1,n]$ | -- |
| $T_{\mathrm{aam}}$ | $F_{\mathrm{aam}}(t)=EXP(\sum_{k\in H}\alpha_{\mathrm{m}k})$ | -- | - | - | - |

Finally, MTTF of VM-based MSFC consisting of $m$ ASFs executed in $n$ AVMMs is calculated according to Equation (18).

$$\mathbf{MTTF} = \sum_{i=0}^{2m+n} V'_{S_i'} h'_{S_i'} \quad (18)$$

# 5 Experiment Results

In this section, we first perform simulation to verify the approximate accuracy of the proposed model and the derived formulas, making our subsequent experimental results more convincing. Then, we conduct sensitivity analysis experiments and numerical experiments to analyze the effects of system parameters on the dependability of VM-based MSFC.

## 5.1 Experimental Configuration

As described in Section 3, our model assumes that the aging times follow exponential distribution, and the recovery times (the holding time of the component from U/R/F state to P state) and failure times follow general distribution. Therefore, in the experiments, we set the aging times and recovery times to follow the exponential distribution, and set the failure times to follow the hypoexponential distribution. TABLE V shows the types of CDFs and the default parameter settings used in the experiments. Here, '--' denotes the setting of the variables depends on the other variables. Some parameters are set according to [11]. We use MAPLE [32] to conduct simulation and numerical analysis experiments. Note that the simulation and numerical analysis experiments can be implemented in any programming language.

## 5.2 Verification of Model and Metric Formula

Fig. 4 and Fig. 5 show the comparison of simulation and numerical results of the steady-state availability and MTTF, respectively. In these figures, 'ana' and 'sim' denote the numerical results and simulation results, respectively. As shown in Fig. 4 and Fig. 5, the numerical results are very close to the corresponding simulation results, which demonstrates the approximate accuracy of our model and formulas.

## 5.3 Sensitivity Analysis

This section analyzes the sensitivity of steady-state availability and MTTF with regard to different parameters when there are four ASFs, four AVMs and two AVMMs in the system. We calculate the sensitivity of metrics by applying Equation (19) [11].

$$SS_{\sigma}(\gamma) = \frac{\partial \gamma}{\partial \sigma}\left(\frac{\sigma}{\gamma}\right) \quad (19)$$

Here, $\gamma$ is a given metric whose sensitivity depends on the parameter $\sigma$. The sensitivity of each metric with respect to different system parameters is given in TABLE VI. We can observe:

- The steady-state availability and MTTF of VM-based MSFC increases with the increasing aging and failure times. However, the steady-state availability increases as the recovery times decrease. MTTF increases as the failover and migration times decreases. Moreover, the recovery times, except for the failover and migration times, are independent of MTTF.
- Compared with other parameters, the aging time has the greatest impact on the steady-state availability and MTTF.

These experimental results can help service providers identify bottlenecks that affect the improvement of the availability and MTTF of VM-based MSFC.

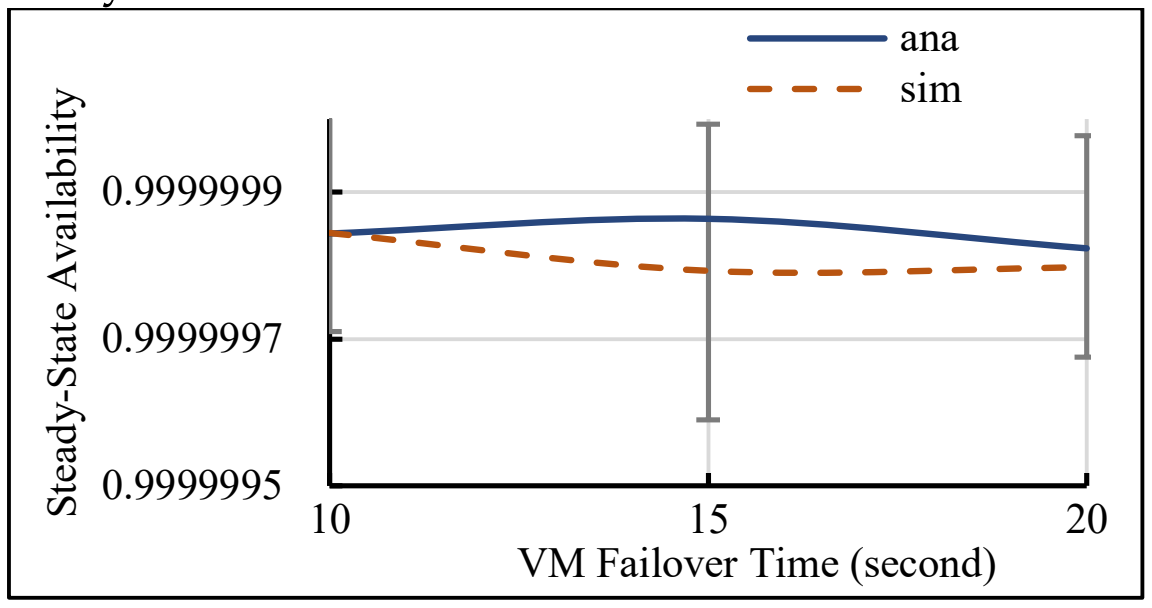

**Fig. 4** Comparison of numerical and simulation results for steady-state availability

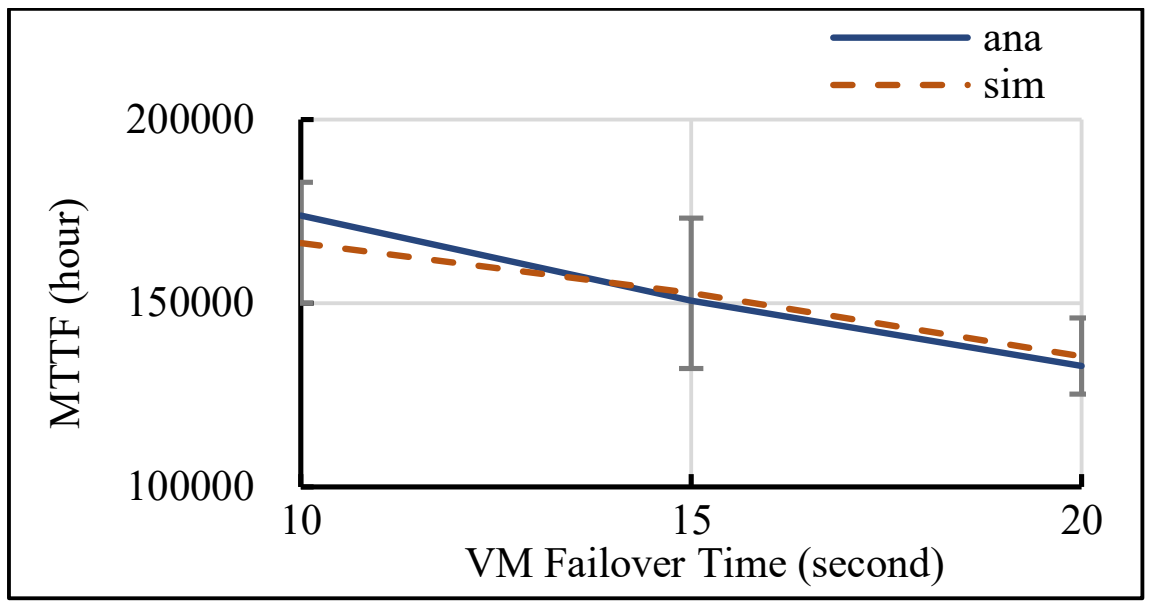

**Fig. 5** Comparison of numerical and simulation results for MTTF

TABLE VI
The Sensitivity of the Steady-State Availability and MTTF

| Variable | Steady-State Availability | MTTF |
|---|---|---|
| $\alpha_{s1}$ | -7.27e-08 | -9.04e-01 |
| $\alpha_{v1}$ | -6.77e-08 | -7.38e-01 |
| $\alpha_{m1}$ | -5.28e-08 | -4.20e-01 |
| $\beta_{s1}$ | 2.10e-08 | 3.05e-01 |
| $\beta_{v1}$ | 3.12e-08 | 3.71e-01 |
| $\beta_{m1}$ | 3.57e-08 | 2.90e-01 |
| $\lambda_{s11}$ | -3.23e-10 | -4.79e-06 |
| $\lambda_{s12}$ | -4.41e-10 | -6.52e-06 |
| $\lambda_{v11}$ | -1.10e-10 | -1.63e-06 |
| $\lambda_{v12}$ | -1.21e-10 | -1.78e-06 |
| $\lambda_{m11}$ | -4.26e-11 | -6.23e-07 |
| $\lambda_{m12}$ | -4.34e-11 | -6.34e-07 |
| $\gamma_{e1}$ | 9.43e-10 | - |
| $\gamma_{s1}$ | 1.48e-09 | - |
| $\gamma_{v1}$ | 5.39e-09 | - |
| $\gamma_{m1}$ | 1.09e-08 | - |
| $\delta_s$ | 5.94e-09 | - |
| $\delta_v$ | 2.87e-08 | - |
| $\delta_m$ | 3.68e-08 | - |
| $\varepsilon$ | 4.77e-10 | - |

## 5.4 Effect of Type of Distribution Function on Steady-State Availability and MTTF

This section gives the steady-state availability and MTTF of VM-based MSFC under different recovery time distribution functions and failure time distribution functions.

We compare system steady-state availability and MTTF for three different distribution function types. First, we assume that the failure times follow the hypoexponential distribution and the recovery times follow the deterministic distribution, represented as 'F-HYPO_R-Deter' in Fig. 6 and Fig. 7. Second, it is assumed that the failure times follow the hypoexponential distribution and the recovery times follow the exponential distribution, represented as 'F-HYPO_R-EXP' in figures. Third, we assume that the failure times follow exponential distribution and the recovery times follow exponential distribution, represented as 'F-EXP_R-EXP' in figures. Fourth, we assume that the failure times follow exponential distribution and the recovery times follow deterministic distribution, represented as 'F-EXP_R-Deter' in figures. Note that in each set of experiments using different failure and recovery time distribution functions, the mean value of each variable is constant.

We can observe:

- The distribution function followed by the failure time has a significant impact on the numerical results. There is a significant difference between the numerical results under exponential failure time and those under non-exponential failure time. Fig. 6 shows that the steady-state availability under the exponential failure time is far different from that under non-exponential failure time. Similarly, it can be seen from Fig. 7 that the difference between MTTF under exponential failure time and MTTF under non-exponential failure time is about $4\times10^4$. This can be explained by the fact that the failure rates of exponential and non-exponential distributions are different.
- The distribution function followed by the recovery time has a minor impact on the numerical results. Fig. 6 and Fig. 7 show that the numerical results under the recovery times following exponential distribution are close to those under the recovery times following non-exponential distribution.

We can conclude that the distribution function type of failure time has a significant effect on each metric.

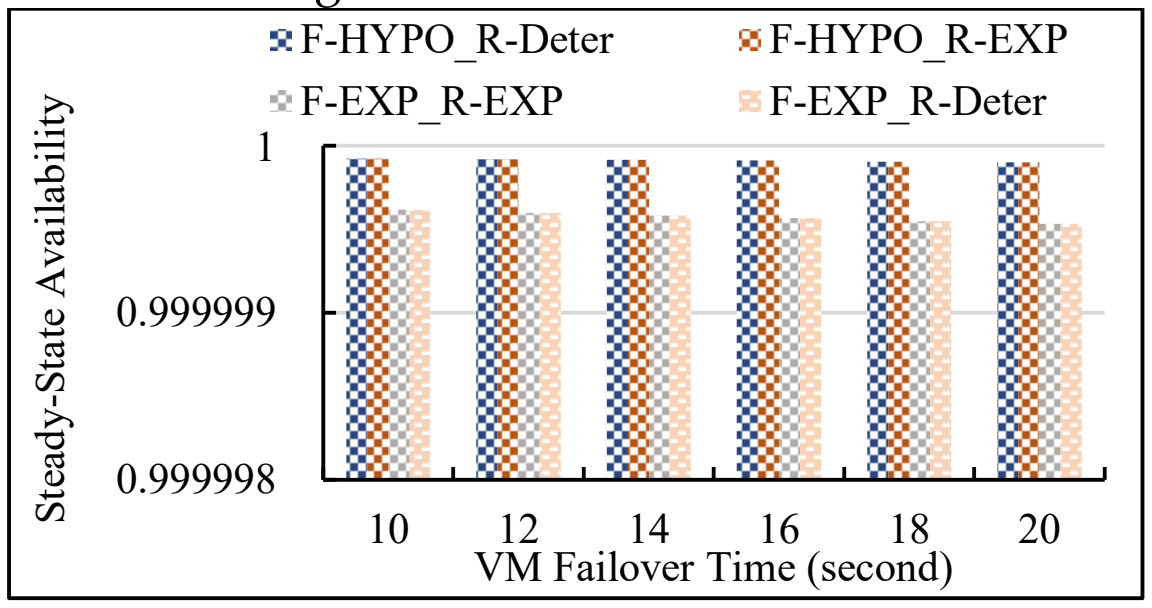

**Fig. 6** Steady-state availability under different failure and recovery time distribution functions

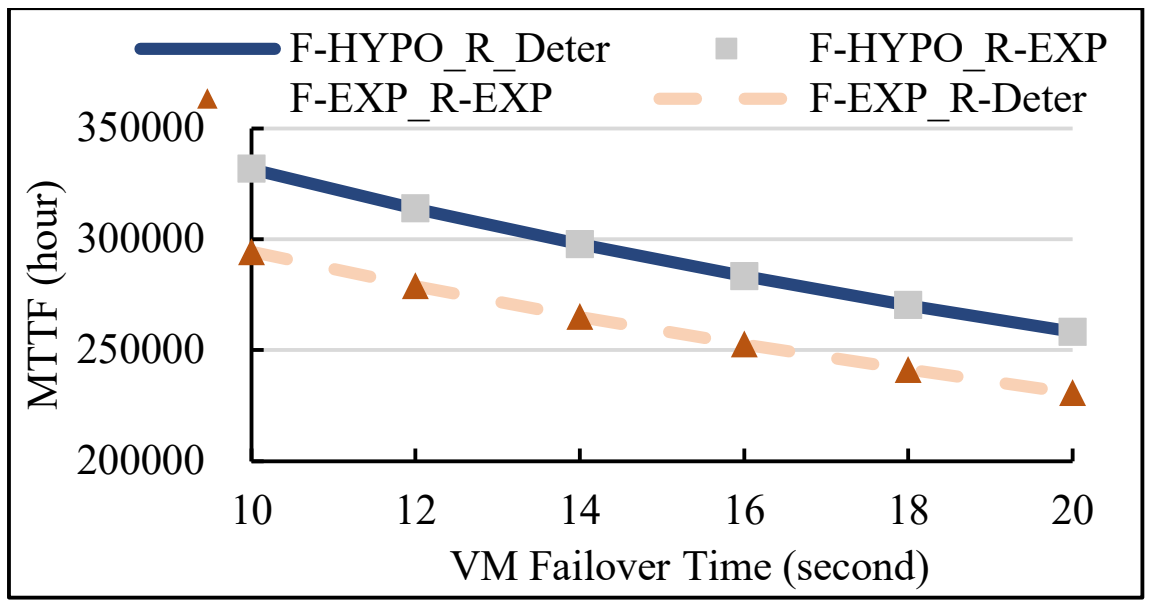

**Fig. 7** MTTF under different failure and recovery time distribution functions

## 5.5 Effect of Failover Rate and Migration Rate on Steady-State Availability and MTTF

This section shows the numerical results of the steady-state availability and MTTF of VM-based MSFC under different VM failover rates and VM migration rates. Fig. 8 and Fig. 9 show the numerical results when VM-based MSFC consisting of four ASFs executed in two AVMMs, respectively. VM failover rate denotes the inverse of VM failover time ($T_{rvi}$, $1 \le i \le 4$) and VM migration rate denotes the inverse of VM migration time ($T_{rmj}$, $1 \le j \le 2$). We can observe from Fig. 8 that the steady-state availability increases with increasing failover and migration rates. The same results can be observed in Fig. 9. It can be explained that the increase in failover time and migration time leads to an increase in the probability of system failure.

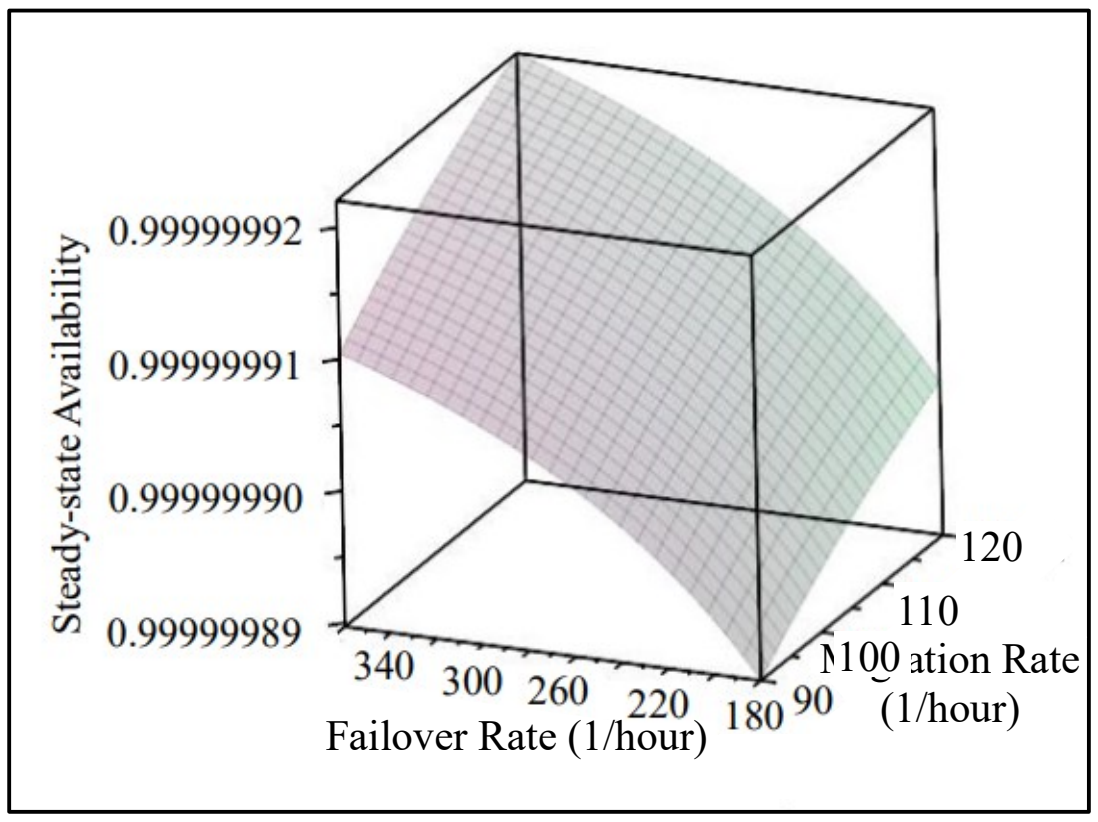

**Fig. 8** Steady-state availability under different failover and migration rates

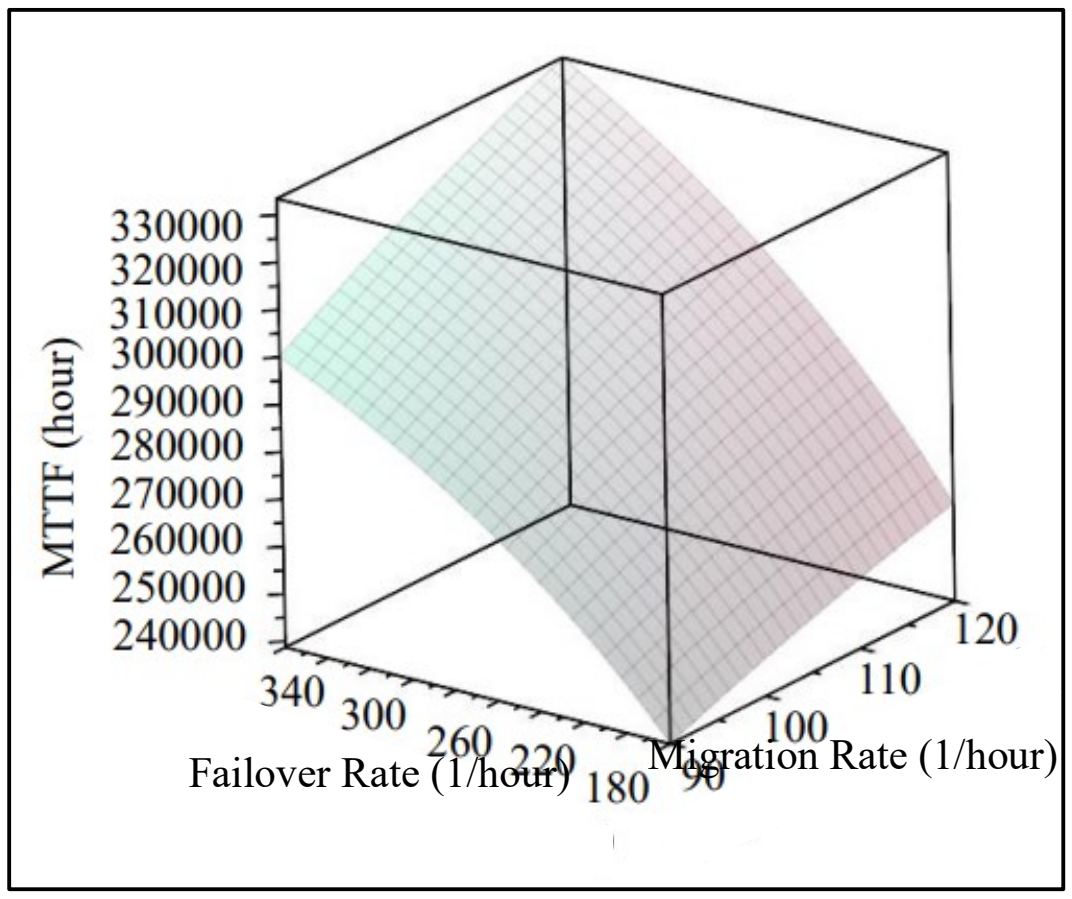

**Fig. 9** MTTF under different failover and migration rates

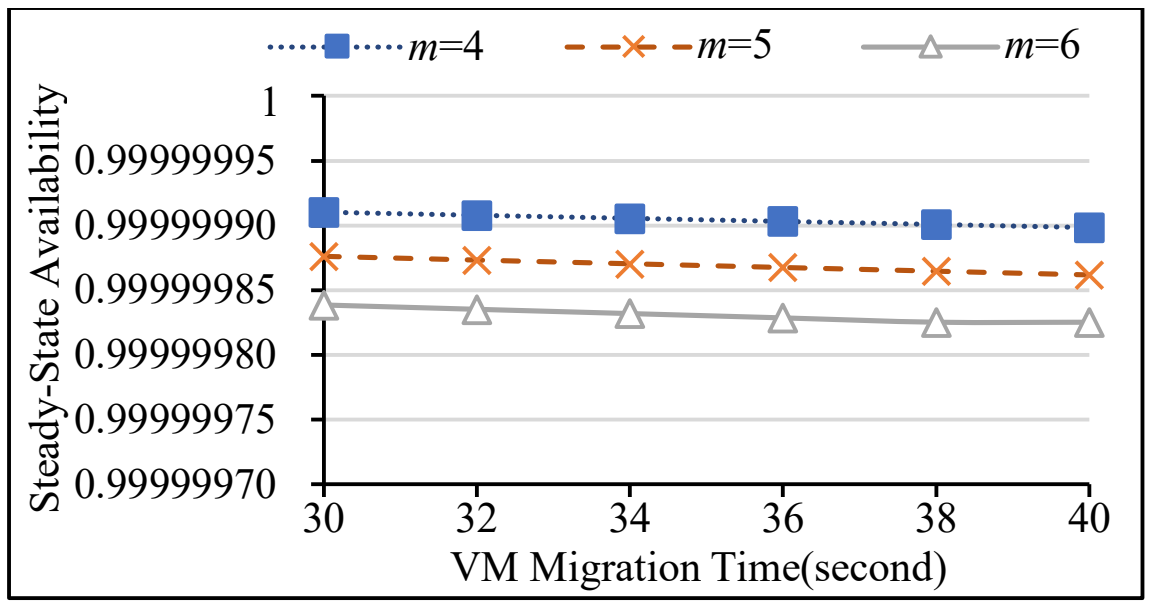

**Fig. 10** Steady-state availability under different number of ASFs and AVMs

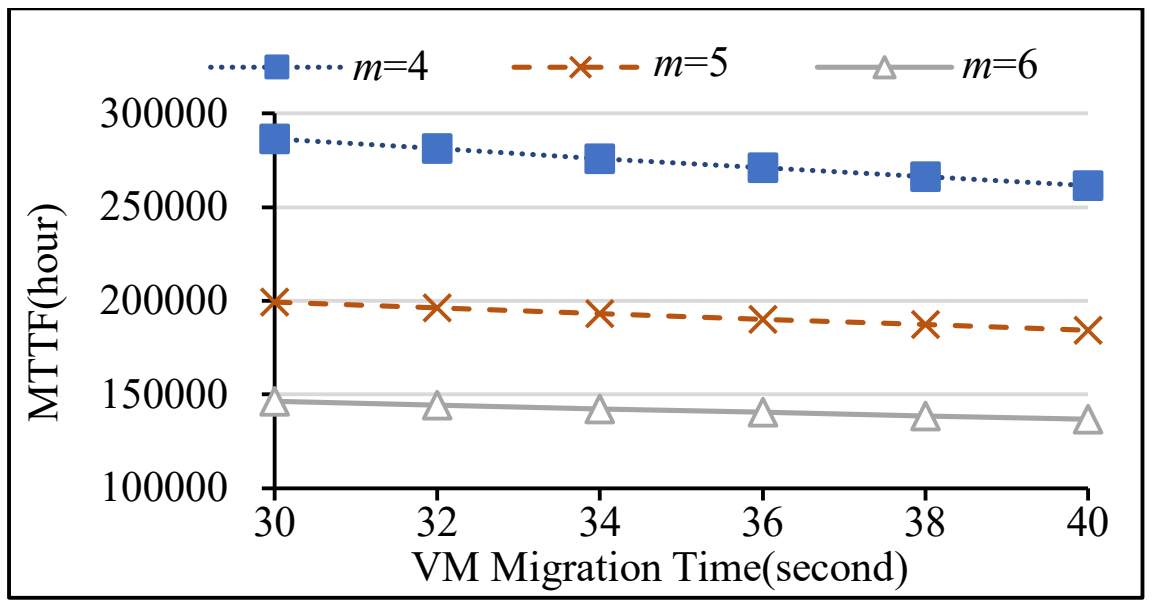

**Fig. 11** MTTF under different number of ASFs and AVMs

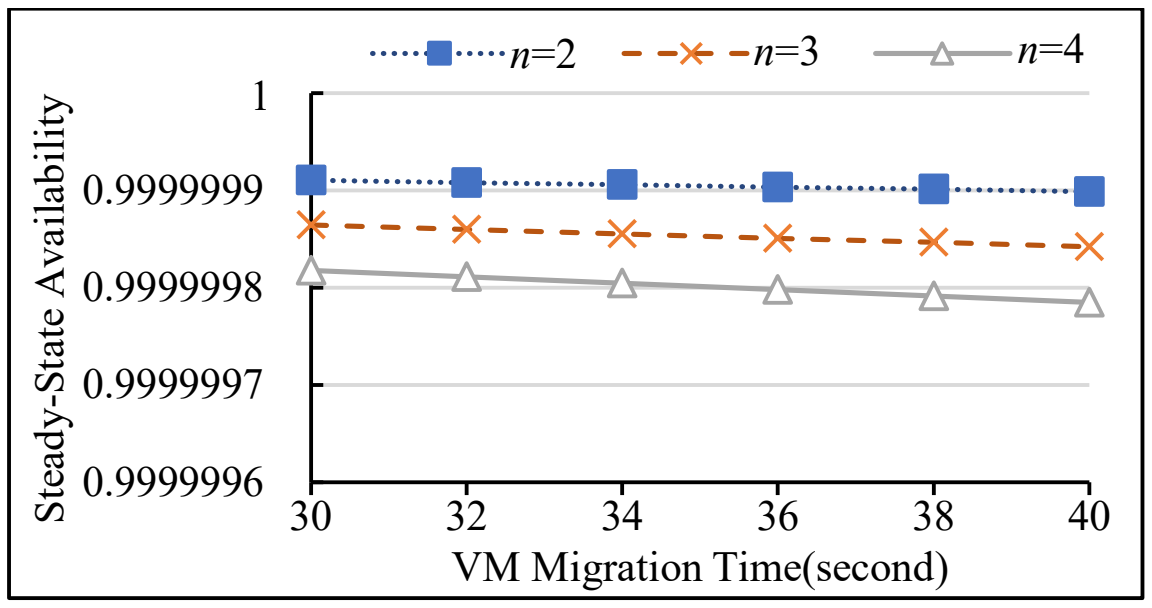

**Fig. 12** Steady-state availability under different number of AVMMs

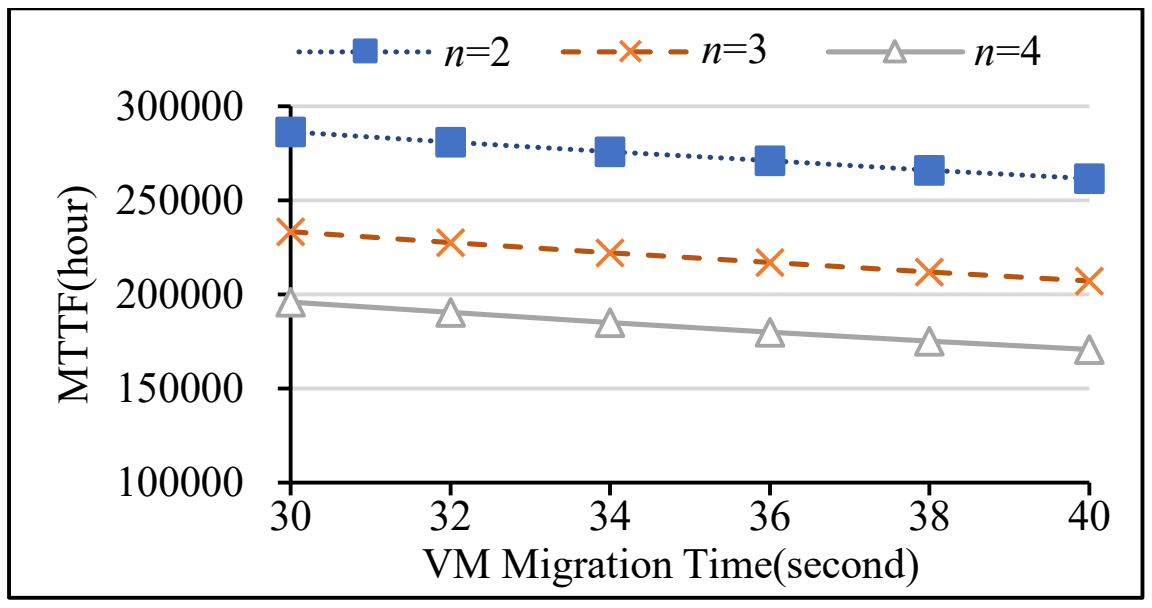

**Fig. 13** MTTF under different number of AVMMs

### 5.6 Effect of Number of Components on Steady-state Availability and MTTF

This section describes the impact of the number of components on the steady-state availability and MTTF. $m$ and $n$ denote the number of ASFs/AVMs and AVMMs, respectively. Without loss of generality, we take $m = 4, 5, 6$ and $n = 2, 3, 4$ as examples to analyze the influence of the number of ASFs and AVMs on both the steady-state availability and MTTF. Fig. 10-13 show the results. We observe that both steady-state availability and MTTF decrease as the number of components increases. It can be explained by the fact that an increase in the number of ASFs, AVMs and VMMs leads to an increase in the number of components that may fail, and causes the system to be in a failure state for a longer period of time.

## 6 Conclusions

In this paper, we propose an SMP model to capture the behaviors of a VM-based MSFC system from software aging to recovery under the condition that the failure times and recovery times are generally distributed. We then derive the formulas for calculating the steady-state availability and MTTF of VM-based MSFC consisting of $m$ ASFs executed in $n$ VMMs. Here, $m$ and $n$ are any interger. After verifying the approximate accuracy of the proposed model and the metric formulas through simulation experiments, we carry out numerical analysis experiments to quantitatively analyze the impact of system parameters on VM-based MSFC dependability. The experimental results show that the SF aging time is an important parameter for VM-based MSFC dependability and the type of failure time distribution has a significant impact on dependability metrics.

This paper assumes that the VM-based MSFC has enough backup resources. In the future, we will investigate the impact of the abnormal behaviors of backup resources on the effectiveness of the rejuvenation techniques.